\documentclass[11pt]{article}

\usepackage{latexsym}
\usepackage{amsfonts}
\usepackage{amsmath}

\newtheorem{theorem}{Theorem}
\newtheorem{definition}[theorem]{Definition}
\newtheorem{lemma}[theorem]{Lemma}
\newtheorem{corollary}[theorem]{Corollary}
\newtheorem{proposition}[theorem]{Proposition}
\newtheorem{remark}[theorem]{Remark}

\numberwithin{theorem}{section}

\newcommand{\R}{I\!\!R}
\newcommand{\N}{I\!\!N}
\newcommand{\Z}{I\!\!Z}
\newcommand{\esssup}{\rm{esssup}}

\newcommand{\essinf}{\rm{essinf}}
\newcommand{\INTER}{\mathop{\rm {\cap}}\limits}

\begin{document}

\title{Time Consistent  Dynamic Limit Order Books\\  Calibrated on Options}

\author{Jocelyne Bion-Nadal \\ 
CMAP (UMR CNRS 7641)  \\
Ecole Polytechnique  
F-91128 Palaiseau cedex \\
bionnada@cmapx.polytechnique.fr}

\date{ }

\maketitle

\begin{abstract}
 In an incomplete financial market, the axiomatic of Time Consistent Pricing Procedure (TCPP), recently introduced,  is used to assign to any financial asset a dynamic limit order book, taking into account both the dynamics of basic assets and the limit order books for options.\\ 
  Kreps-Yan  fundamental theorem is extended to that context. A characterization of TCPP calibrated on options is given in terms of their dual representation. In case of perfectly liquid options, these options can be used  as the basic assets to hedge dynamically. A generic family of TCPP calibrated on option prices is constructed, from  c\`adl\`ag BMO martingales. 
\end{abstract}
\section{Introduction} 
\label{sec:intro}

 The problem of  dynamic pricing is  the problem of extending a function that gives the prices of marketed financial instruments to a larger class of financial instruments. 
The usual way of dynamic pricing in financial mathematics is to start with a (No Free Lunch) dynamic model for the stock prices and to use the theory of portfolios constructed from these basic assets to price the other financial instruments. 
 The first step along these lines was made by Black Scholes and Merton. 
In a complete market the dynamic price of any financial instrument $X$ is then equal to the dynamic price of the replicating portfolio. As pointed out by Avellaneda and Paras \cite{ALP} and \cite{AP}, the market prices of options give important informations on the volatility. Therefore the prices constructed from the theory   
associated with options have to be compatible with their observed bid and ask prices. If the constructed prices do not lie within the interval defined by the observed bid and ask prices, this means that the choosen dynamics for the basic assets induce an arbitrage in the financial market. For example the Black Scholes model with constant volatility is not compatible with the call and put options prices. This is the volatility smile effect.\\
In a Brownian setting,  the notion of implied volatility has been introduced, inverting the Black Scholes formula for the options prices. Then a wide litterature has been developped trying to modelize the implied volatility.  Necessary conditions for the resulting model to be arbitrage free have been given by Sch\"onbucher \cite{S}. However  there exists no dynamic model for implied volatility leading to arbitrage free prices. 
Other approaches have been developped in order to produce, in a arbitrage free way, dynamic prices consistant with observed prices for options. The local volatility model introduced by  Dupire \cite{Du} is an arbitrage free dynamic model of one stock, in a Brownian setting,  assuming a particular shape for the volatility. It assumes also that one observes in the market the prices of call options on this asset for all strikes and all maturity dates and furthermore that the corresponding function is very smooth (in particular of $C^2$ class in the strike). This leads to a non robust model. In addition  this model is a complete model for one stock which is not compatible with some observations in the market.
Other approaches have been introduced recently in order to price dynamically in a arbitrage free way, taking into account the observed prices for options.
Jacod and Protter \cite{JP} and Schweizer and Wissel \cite{SW1} assume that only options with one fixed payoff function but all maturities are traded. Schweizer and Wissel \cite{SW2} consider also the case where call options with one fixed maturity but all strikes are traded. In both cases the dynamics of the stock and of the options are modelized simultaneously in a arbitrage free way.
However in real financial markets options of various kind with various strikes and various maturities are traded. Only a finite number of options are traded and not a continuum.\\ 
Furthermore the options are not all perfectly liquid. At some fixed instant only a limit order book is observed for some  options and not a price. For $n$ large enough the ask price of $nX$ is larger than $n$ times the ask price of $X$. This implies that the market is incomplete. 
When  the model for the stock prices is  not complete, there are several equivalent local martingale measures for the stock prices, or equivalently financial assets are not perfectly replicated by portfolios in the basic assets. 
Thus a natural way of assigning a dynamic ask price to a financial asset $X$ defined at time $T$ (for example an option of maturity date $T$),  using the theory of portfolios, is to consider portfolios in the basic assets dominating at time $T$ this asset $X$. This leads to the  super-replication price, originally studied by El Karoui and Quenez \cite{EKQ}, which is the minimal price of portfolios in the basic assets dominating $X$. The super-replication price is sublinear.  The dynamic super-replication price is equal to $\esssup_{Q \in {\cal Q}} E_Q(X|{\cal F}_t)$, where ${\cal Q}$ is the set of all equivalent local martingale measures for the stock prices. 
However for many models this super-replication price is too high. It doesn't lie within the interval defined from the observed bid and ask prices associated with the option. Notice also that in case of linear or sublinear ask prices, the  ask price associated with $nX$ ($n \geq 0$) for any financial asset $X$ is equal to $n$ times the  ask price associated with $X$. Linear or sublinear prices don't take into account the liquidity risk. On the contrary 
the observation of limit order books   leads to the conclusion that for any traded asset $Y$ the ask price of $nY$ is a convex,  not sublinear, function of $n$.\\ The context of the present paper   is that of an incomplete and illiquid market. We construct  a dynamic pricing theory taking into account both the dynamics of basic assets and the limit order books of options on these assets. This is done making use of the theory of No Free Lunch TCPP introduced in \cite{BN05}. 
 We  consider a reference family  composed of two kinds of assets: the basic assets $(S^k)_{0 \leq k \leq d+1}$ for which the dynamic process is assumed to be known , and the assets $(Y^l)_{1 \leq l \leq d})$ (for examples options) which are only  revealed at their maturity date (the stopping time $\tau_l$) and for which one observes a limit order book  at time $0$. One of the basic asset  $S^0$ is assumed to be strictly positive and is taken as num\'eraire. \\
 The first question we address is the question of non existence of  arbitrage for the reference family $((S^k)_{0 \leq k \leq d+1},(Y^l)_{1 \leq l \leq d})$ and the observed limit order books associated with the assets $Y^l$. 
 We extend to that context the notion of No Free Lunch, replacing the usual notion of dynamic strategy with respect to the basic assets $(S^k)$ by the sum of a dynamic strategy with respect to the basic assets $(S^k)$ and of a static strategy with respect to the options $Y^l$. 
We prove the following generalization of  Kreps-Yan Theorem: there is No Free Lunch with respect to the reference family  if and only  there is an equivalent local martingale measure $Q$ for the process $(S^k)_k$ such that, for every $l$, and any $n \geq 0$, $C_{bid}(nY^l) \leq E_Q(nY^l) \leq C_{ask}(nY^l)$.
 The conditional expectation with respect to $Q$ provides then a linear pricing procedure calibrated on the reference family.  However as mentioned above, in order to take into account the liquidity risk,  we do not want to restrict to linear nor sublinear pricing procedures.
The theory of No Free Lunch TCPP  takes into account the liquidity risk and allows for the  construction of dynamic limit order books associated with any financial instrument. This construction is done in a arbitrage free way and consistently in time.\\
The second main result is the characterization of TCPP  calibrated on the 
reference family $((S^k)_{0 \leq k \leq d},(Y^l)_{1 \leq l \leq p})$  and the limit order books  in terms of their dual representation. We also study the supply curves for such TCPP.\\ 
The third result concerns the study of the hedge in the case   where both the basic assets $S^k$ and the options $Y^l$ are assumed to be very liquid. In that case we prove that the options can be used to hedge dynamically as well as the basic assets. \\
The last important result of the paper is the generic construction of a family of convex No Free Lunch TCPP calibrated on option prices. We prove the existence of a non sublinear TCPP calibrated on option prices belonging to the class first introduced in \cite{BN03}, making use of the theory of right continuous BMO martingales, as soon as the reference family satisfies the robust No Free Lunch condition. This construction is made in a very general setting of locally bounded stochastic processes, for which  jumps are allowed. \\

\section{First Fundamental Theorem}
\label{sec:fundamental Theorem} 
\subsection{The economic model}
\label{sec:model} 
 We work with  a filtered probability space $(\Omega,{\cal F}_{\infty},({\cal F}_t)_{ t \in \R^+}, P)$ throughout this paper. The 
filtration $({\cal F}_t)_{ t \in \R^+}$ satisfies the usual assumptions of right continuity and completeness and ${\cal F}_0$ is assumed to be the $\sigma$-algebra generated by the $P$ null sets of ${\cal F}_{\infty}$.  We assume that the time horizon is infinite, which is the most general case. Indeed if the time horizon is finite equal to $T$ we define ${\cal F}_s={\cal F}_T$ for every $s \geq T$.\\ 
The usual way of dynamic pricing is to start with some reference assets for which the dynamics is assumed to be known and to  construct a dynamic pricing procedure extending the dynamics of these reference assets. In order to use more information from the market, we want to take also into account the limit order books associated with some options. \\ 
 Therefore the reference family will be composed of two kinds of assets: As usual, we consider that there are some basic assets $(S^k)_{0 \leq k \leq d}$ for which we have a good idea of the evolution of their dynamics and we want to take into account all these dynamics. From a newer point of view there are also assets $(Y^l)_{1 \leq l \leq p}$ as options  of various maturity dates on one or several of the basic assets on which there are a lot of transactions, so that it is meaningful to take into account the corresponding limit order books  observed in the market. Notice that  even if one  knows  the dynamics of the underlying assets, one  doesn't know, in an incomplete market,  the dynamics of  options. The option   is revealed at time $\tau_l$ which is the maturity date of the option. The value at time $\tau_l$ of this option is therefore modeled by a ${\cal F}_{\tau_l}$ measurable function $Y^l$.
 We assume that in the market at time $0$,  a limit order book  is observed for each of the options $(Y^l)_{1 \leq l \leq p}$.\\
 We assume that $S^0$ is always positive, and  we can take it as num\'eraire. So from now on,
$(S^0)_t=1 \; \forall \; t \in \R^+$,  $S_t=(S^k_t)_{1 \leq k \leq d}$ models the discounted price process of $d$ risky assets, and  $Y^l$ the discounted prices of options (at time $\tau_l$).  $S$ is assumed to be a locally bounded  stochastic 
process with a.s. c\`adl\`ag trajectories.\\ 
 For any stopping time $\tau$, denote ${\cal F}_{\tau}$ the $\sigma$-algebra defined by\\
${\cal F}_{\tau}=\{A \in \ {\cal F}_{\infty}|\forall \; t \in \R^+\; A \INTER \{\tau \leq t \} \in {\cal F}_t\}$. 
Denote $L^{\infty}(\Omega,{\cal F}_{\tau},P)$ the Banach algebra of essentially bounded  real valued ${\cal F}_{\tau}$ 
measurable functions. We will always identify an essentially bounded ${\cal F}_{\tau}$ measurable function with its class in $L^{\infty}(\Omega,{\cal F}_{\tau},P)$.\\
The aim of this section is to define a notion of no arbitrage extending the usual one and to prove in this new context  a first fundamental theorem generalizing the Kreps Yan theorem. 
Before that we want to point out some properties of the limit order books.
\subsection{Limit order books}
\label{sec:book}
 Let $Y^l$ be a traded financial asset.  One assumes that at time $t_0$ one observes a limit order book  associated with the asset $Y^l$. The limit order book takes into account only the non executed  orders at time $t_0$.  One assumes that all the non executed orders on the asset $Y^l$ are written in the following tabular.

\begin{center}
\begin{tabular}{|c|c|p{2cm}|c|c|}
\cline{1-2} \cline{4-5}
\multicolumn{2}{|c|}{Bid } &  $ $ & \multicolumn{2}{c|}{Ask }\\ \cline{1-2} \cline{4-5}
quantity   & limit  & $ $ &  limit & quantity \\ \cline{1-2} \cline{4-5}
$M_1$ &  $ C^{1}_{bid}$ &  $ $ &$ C^{1}_{ask}$ & $N_1$\\ \cline{1-2} \cline{4-5}
$M_2$ & $C^{2}_{bid}$ &  $ $ &$C^{2}_{ask}$ &$ N_2$\\ \cline{1-2} \cline{4-5}
$...$ & $...$ & $ $ &  $... $ &$...$\\ \cline{1-2} \cline{4-5}
$M_p$ & $C^{p}_{bid}$ &  $ $ &$C^{q}_{ask}$ & $N_q$\\ \cline{1-2} \cline{4-5}

\end{tabular}
\end{center}
with \begin{equation}C^{p}_{bid} <...< C^{1}_{bid} < C^{1}_{ask}<... < C^{q}_{ask}
\label{eqba}
\end{equation}
If there is also a transaction at time $t_0$ on the asset $Y^l$, we denote $C^0$ the price of the transaction and $N_0=M_0$ the number of shares exchanged at time $t_0$ (if there is no transaction on $Y^l$ at time $t_0$, $N_0=M_0=0$).  Necessarily, 
\begin{equation}
 C^{1}_{bid} \leq C^0 \leq  C^{1}_{ask} 
\label{eqba2}
\end{equation}
Taking into account the limit order book, one can canonically associate to any positive integer $n \leq \sum_{0 \leq i \leq q} N_i=N^l$  the ask price $C_{ask}(nY^l)$ defined as follows:
Let $j \leq q$ be such that $\sum_{0 \leq i \leq j-1} N_i \leq n <\sum_{0 \leq i \leq j} N_i$. Define $$C_{ask}(nY)=\sum_{0 \leq i \leq j-1} N_i C^i_{ask}+(n-\sum_{0 \leq i \leq j-1} N_i)C^{j}_{ask}$$ if $0 \leq n \leq N_0$, $C_{ask}(nY)=nC^0$.
 The bid price associated with $nY$ for $n \leq \sum_{0 \leq i \leq p} M_i=M^l$ is defined in a similar way.\\
Notice that it is easy to verify from the definition of $C_{ask}(nY)$ and the relations (\ref{eqba}) and (\ref{eqba2}) that the map $n \in \N \rightarrow C_{ask}(nY)$ is convex and the map $n \in \N \rightarrow C_{bid}(nY)$ is concave. In particular $n \in \N \rightarrow \frac{C_{ask}(nY)}{n}$ is increasing  and $n \in \N \rightarrow \frac{C_{bid}(nY)}{n}$ is decreasing. Also for any $n, m\;\; \frac{C_{bid}(mY)}{m} \leq \frac{C_{ask}(nY)}{n}$.
\subsection{Fundamental Theorem}
\label{sec:fund. Thm} 
The first step is to define the notion of admissible simple strategy in this new setting. The investor can use two kinds of assets. The basic assets  $S^k$ for which the dynamics are assumed to be known, therefore an investor can trade dynamically using the $S^k$. He can also invest in the assets $Y^l$ but these assets are only known at their maturity date $\tau_l$ and not at any intermediate date, therefore we restrict to static investments on $Y^l$ between the dates $0$ and $\tau_l$. From the observation of the limit order book  associated with $Y^l$ at time $0$ we associate, as in the previous Section \ref{sec:book}, to any $n \leq N^l$ an ask price $C_{ask}(nY^l)$ and to any $n \leq M^l$ a bid price $C_{bid}(nY^l)$.

\begin{definition}
An admissible simple strategy  with respect to the reference assets $((S^k)_{0 \leq k \leq d},(Y^l)_{1 \leq l \leq p})$ is the sum  of a dynamic simple strategy $H$ with respect to the process $(S^k)$ and of a static strategy with respect to the random variables $(Y^l)$.\\ $H=\sum_{i=1}^{n} h_i {\cal X}_{]\sigma_{i-1},\sigma_i]}$, where $0 =\sigma_0 \leq \sigma_1 \leq ...\leq \sigma_n$ are finite stopping times and $h_i$ are essentially bounded $\R^d$ valued ${\cal F}_{\sigma_{i-1}}$ measurable functions and the stopped process $(S^k)^{{\sigma}_n}$ is uniformly bounded.
\end{definition}

Define now the convex set of  contingent claims available at  zero or negative price, using admissible simple strategies, taking into account  the fact that for the random variables $Y^l$, one observes a limit order book. In all the following the limit order book  observed for $Y^l$ will be denoted 
$(C_{bid}(mY^l), C_{ask}(nY^l))$. This means: \\
$(C_{bid}(mY^l)_{0 \leq m \leq M^l}, C_{ask}(nY^l)_{0 \leq n \leq N^l})$.

\begin{definition} The convex set  of portfolios available at zero cost is:
$$K=\{\sum_{i=1}^{n} \sum_{k=1}^{d} (h^k_i) (S^k_{\sigma_i}-S^k_{\sigma_{i-1}})+
\sum_{l=1}^{p}(\gamma^l-\beta^l)Y^l +(\gamma^0-\beta^0)\;; (h^k)_i \in L^{\infty}({\cal F}_{{\sigma}_{i-1}}),$$
$$ \beta^l, \gamma^l \in \N \; \beta^l \leq N^l,\; \gamma^l \leq M^l \; |\;\sum_{l=1}^{p}( C_{ask}(\gamma^l Y^l)- C_{bid}(\beta^lY^l)+(\gamma^0-\beta^0) \leq 0 \}.$$
\end{definition}
Notice that adding the condition: for any $l$, either $\gamma^l$ or $\beta^l$ is equal to $0$ in the definition of $K$ would not change the set $K$.
An element of $K$ is  the sum of a static portfolio in the options $Y^l$ corresponding to $\gamma^l$ long position in $Y^l$ and $\beta^l$ short position in $Y^l$ and of  a dynamic portfolio in the assets $S^k$ available at price $0$. The convexity of $K$ follows from  the convexity (resp concavity) of the map $\gamma \rightarrow C_{ask}(\gamma Y^l)$ (resp $\beta  \rightarrow C_{bid}(\beta Y^l)$). Denote $\tilde K$ the set of portfolios dominated by an element of $K$, $\tilde K =K-L^{\infty}_+$.\\
 In this setting, we say that there is No Arbitrage if there is no non trivial   non negative  attainable claim ot zero cost, i.e. $K \INTER L^{\infty}_+(\Omega,{\cal F},P)=\{0\}$. Notice that this condition is equivalent to   $C \INTER L^{\infty}_+(\Omega,{\cal F},P)=\{0\}$, where $C$ is the cone generated by $\tilde K$.\\
We prove now a theorem generalizing the Kreps Yan theorem to that context. As in the usual setting, the notion of No Arbitrage is not sufficient, we have to pass to the notion of No Free Lunch.

\begin{definition}

The reference family $((S^k)_{0 \leq k \leq d},(Y^l)_{1 \leq l \leq d})$ satisfies the No Free Lunch condition with respect to the limit order books  $C_{bid}(mY^l), C_{ask}(nY^l)$ if the closure $\overline C$ of $C$ with respect to the weak* topology of $L^{\infty}(\Omega, {\cal F},P)$  satisfies $\overline C \INTER L^{\infty}_+(\Omega,{\cal F},P)=\{0\}$.
\end{definition}
 First fundamental theorem generalizing Kreps-Yan theorem:
\begin{theorem}
The following conditions are equivalent:
\begin{itemize}
\item[i)] The reference family  $((S^k)_{0 \leq k \leq d},(Y^l)_{1 \leq l \leq d})$ satisfies the No Free Lunch condition with respect to the limit order books  $C_{bid}(mY^l), C_{ask}(nY^l)$.
\item[ii)]
 There is an equivalent local martingale measure  $R$ for  $(S^k)_{0 \leq k \leq d}$ such that for any 
$l \in {1,...,p}$, for $ m \leq M^l\;\;C_{bid}(mY^l) \leq E_R(mY^l)$ and for $n \leq N^l\;\;E_R(nY^l)\leq C_{ask}(nY^l)$.
\end{itemize}
\label{thm1}
\end{theorem}
We will give  a more complete version of this Theorem in Section \ref{sec:EV} after having discussed the notion of  TCPP calibrated on option prices. The proof is in Appendix \ref{secEFFT}.
\section{TCPP calibrated on options}
\label{sec:admissibility}  
\subsection{TCPP calibrated on a reference family}
Recall briefly the definition of TCPP (Time Consistent Dynamic Pricing Procedure) , that we have introduced in \cite{BN05}  in order to assign to any financial product a dynamic limit order book  in a financial market with transaction costs and liquidity risk. 
Other definitions close to the following one can be found in Peng \cite{Peng}, with deterministic times instead of stopping times and in the restrictive context of a Brownian filtration, in Cheridito et al \cite{CDK} in a discrete time setting, and in Kl\"oppel and Schweizer \cite{KS} with just one deterministic time.
 \begin{definition}
Let $(\Omega,{\cal F}_{\infty},({\cal F}_t)_{t \in \R^+},P)$ be a filtered probability space. A TCPP  $(\Pi_{\sigma, \tau})_{0 \leq \sigma \leq \tau}$  (where $\sigma \leq \tau$ are  stopping times) is a 
family of maps $$\Pi_{\sigma, \tau}: 
L^{\infty}({\cal F}_{\tau}) \rightarrow L^{\infty}({\cal F}_{\sigma})$$ satisfying  the properties of  monotonicity, translation invariance,  convexity, normalization, continuity from below and time consistency. \\
For any $X \in L^{\infty}({\cal F}_{\tau})$, the dynamic ask (resp. bid) price process of $X$ is $(\Pi_{\sigma,\tau}(X))_{\sigma}$ (resp. $(-\Pi_{\sigma,\tau}(-X))_{\sigma}$).\\
A TCPP is called sublinear if furthermore $\forall \lambda>0\;\;\forall X \in  L^{\infty}({\cal F}_{\tau}),\;\; \Pi_{\sigma,\tau}(\lambda X)=\lambda \Pi_{\sigma,\tau}(X)$. 
\label{definition1}
\end{definition}
Recall that for any $X \in L^{\infty}({\cal F}_{\tau}), \; -\Pi_{\sigma,\tau}(-X) \leq \Pi_{\sigma,\tau}(X)$
Notice that it follows from time consistency and normalization that for any $\nu \leq \sigma \leq \tau$, $\Pi_{\nu, \sigma}$ is the restriction of $\Pi_{\nu,\tau}$ to $L^{\infty}({\cal F}_{\sigma})$. Remark that a  TCPP assigns to any financial instrument $X \in L^{\infty}({\cal F}_{\infty})$ not only a dynamic bid and ask prices, but also a dynamic limit order book  $(-\Pi_{\sigma,\infty}(-nX), \Pi_{\sigma,\infty}(nX))_{\sigma}$, satisfying at any time, the properties observed for limit order books  in real financial markets  (Section \ref{sec:book}).\\
A TCPP is equal, up to a minus sign, to  a normalized time consistent dynamic risk measure.\\
In this paper we  restrict our attention to No Free Lunch TCPP. For the definition and the general study of No Free Lunch TCPP  we refer to \cite{BN05}. 
Recall in particular that  any No Free Lunch TCPP has a dual representation in terms of equivalent probability measures of finite penalty:
 \begin{equation}
\forall \; X \in L^{\infty}({\cal F}_{\tau}), \;\;\Pi_{\sigma,\tau}(X)= \esssup_{Q \in  {\cal M}^{1,e}(P)}(E_Q(X|{\cal F}_{\sigma})-\alpha^m_{\sigma,\tau}(Q))
\label{eq_NFL2}    
\end{equation}
where
\begin{equation}
{\cal M}^{1,e}(P)=\{ Q\sim  P\; \mbox{and } \; \alpha^m_{0,\infty}(Q) < \;\infty\} 
\label{eq_2NFL}
\end{equation}
Recall also that from \cite{BN05}, 
the No Free Lunch property implies that the set $ {\cal M}^0$  of probability measures equivalent with $P$ with zero minimal penalty is non empty.
\begin{equation} {\cal M}^0=\{ R   \sim  P, \;\; \alpha^m_{0,\infty}(R)=0 \}
\label{eqzeropen} 
\end{equation}
Recall also that any probability measure $Q \in {\cal M}^{1,e}(P)$ satisfies the cocycle condition (cf \cite{BN05}):
\begin{equation}
\forall \nu \leq \sigma \leq \tau \;\; \alpha^m_{\nu,\tau}(Q)=\alpha^m_{\nu,\sigma}(Q)+E_Q(\alpha^m_{\sigma,\tau}(Q))
\label{eqcoc}
\end{equation}
We have proved in \cite{BN05}, that for any $R \in {\cal M}^0$, the ask price (resp bid price) process associated with any $X \in L^{\infty}({\cal F}_{\infty})$ is then a $R$-supermartingale (resp $R$-submartingale) admitting a  c\`adl\`ag modification.
\begin{remark} For any $R \in {\cal M}^0$, for any stopping times $\sigma \leq \tau$, $\alpha^m_{\sigma,\tau}(R)=0$.
\label{remzeropenalty}
\end{remark}
This is an easy consequence of the non negativity and of the cocycle condition satisfied by the minimal penalty (equation \ref{eqcoc}).
\begin{remark}
 Assume now that  $X$ belongs to $L^0(\Omega, {\cal F}_{\infty},P)$, is no more essentially bounded but is such that  $X^{^-} \in L^{\infty} (\Omega, {\cal F}_{\infty},P)$. $X$ is the increasing limit of a sequence  $(X_n)_{n \in \N}$ of elements  in $L^{\infty} (\Omega, {\cal F}_{\infty},P)$. And therefore using the continuity from below of  the TCPP, for any $\sigma$, $\Pi_{\sigma,\infty}(X)$ is defined as the increasing limit of $\Pi_{\sigma,\infty}(X_n)$.
\end{remark}
We give now the definition of calibration of a TCPP on a reference family, definition extending the notion first introduced in \cite{BN05}.
\begin{definition}
A TCPP $(\Pi_{\sigma, \tau})_{0 \leq \sigma \leq \tau}$ is calibrated on  the 
reference family $((S^k)_{0 \leq k,\leq d},$ $(Y^l)_{1 \leq l \leq p})$  and the limit order books  $(C_{bid}(mY^l),C_{ask}(nY^l))_{1 \leq l \leq p}$ if   
\begin{itemize}
\item it extends the dynamics of  the process  $(S^k)_{0 \leq k \leq d}$, i.e.
for any finite stopping time $\tau$ such that the stopped process $(S^k)^{\tau}$ is uniformly bounded,
\begin{equation}
\forall \; n \in \Z\;\; \forall \; 0 \leq \sigma \leq \tau \;\;\;\;\Pi_{\sigma,\tau}(n S^k_{\tau})= n S^k_{\sigma}
\label{eqSAS}
\end{equation}
\item it is compatible with the limit order books of $Y^l$: $\forall  1 \leq l  \leq p $,
\begin{eqnarray}
\forall m \leq M^l\;\; C_{bid}(mY^l) \leq -\Pi_{0,\tau_l}(-mY^l)\nonumber\\ 
\forall n \leq N^l\;\; \Pi_{0,\tau_l}(nY^l) \leq C_{ask}(nY^l)
\label{eqSAY}
\end{eqnarray}
\end{itemize}
\label{definition2}
\end{definition}
 In the condition (\ref{eqSAS}) we consider only integer multiples of the process $(S^k)_{0 \leq k \leq d}$  because there are the only one that can be traded (however considering the definition with real numbers instead of integers would not affect the results).\\ 
The preceding notion of calibration  assumes that the assets $(S^k)_{0 \leq k \leq d}$ are perfectly liquid.  
This is of course not completely realistic. If we want to take into account the existence of a limit order book  associated with the $S^k$, we have to weaken the preceding condition. This is the subject of the next subsection.
\subsection{Weak calibration for a TCPP}
 In this section we introduce a weaker notion of  calibration on the reference family, taking into account the fact that  the financial assets $(S^k)_{0 \leq k \leq d}$ are not perfectly liquid. We want to construct a dynamic for the limit order books, taking into account both the limit order books observed for the process $(S^k)$ at time $0$, ($C_{bid}(nS^k)$, $C_{ask}(nS^k)$), and the dynamics of $S^k$. 
Thus we introduce the following definition of weak calibration:
\begin{definition}
 A TCPP  $(\Pi_{\sigma, \tau})_{0 \leq \sigma \leq \tau}$ is weakly calibrated on  the reference 
family $((S^k)_{0 \leq k \leq d},(Y^l)_{1 \leq l \leq p})$ and the observed limit order books\\ $(C_{bid}(m S^k),C_{ask}(nS^k))_{1 \leq k \leq d}$, $(C_{bid}(m Y^l),C_{ask}(nY^l))_{1 \leq l \leq p}$ if it satisfies the following conditions:\\ 
1.Weak extension of the process  $(S^k)_{0 \leq k \leq d}$: For any finite stopping time $\tau$ such that the stopped process $(S^k)^{\tau}$ is uniformly bounded, 
$\forall \; 0 \leq k \leq d$  $\forall \; 0 \leq \sigma \leq \tau$
\begin{itemize}
\item[i) ]
 $\forall n \in \N$
 $ -\Pi_{\sigma,\tau}(-n S^k_{\tau}) \leq  nS^k_{\sigma} \leq \Pi_{\sigma,\tau}(n S^k_{\tau})\;\;$
\item[ii) ]
 $ -\Pi_{\sigma,\tau}(- S^k_{\tau})= \Pi_{\sigma,\tau}( S^k_{\tau})= S^k_{\sigma}$
\item[iii) ] compatibility with the limit order books  of $(S^k)_{0 \leq k \leq d}$,
$$\forall  n \leq M(k)\;C_{bid}(nS^k) \leq -\Pi_{0,\tau}(-n S^k_{\tau})$$
$$ \forall  n \leq N(k) \Pi_{0,\tau}(n S^k_{\tau}) \leq C_{ask}(nS^k)$$
\end{itemize}
2. Compatibility with the limit order books of $(Y^l)_{1 \leq l \leq p}$:
 equation (\ref{eqSAY}) of Definition \ref{definition2}.
\label{definition3}
\end{definition}
\begin{remark}
 For a sublinear TCPP there is no difference between  calibration and  weak calibration.
\label{rem2}
\end{remark}
As proved  in the next lemma, if there is  a transaction on  each $S^k$ at time zero,  the condition 1.{\it ii)} is a consequence of the assumptions   1.{\it i) and {iii)}}.
\begin{lemma}
Assume that there is at time zero a transaction on each  $S^k$, i.e.
$C_{bid}(S^k)=C_{ask}(S^k)$. Assume that the pricing procedure satisfies the conditions 1.{\it i)} and {\it iii)} of Definition \ref{definition3}, then it also satisfies condition 1.{\it ii)}.  
\label{lem1}
\end{lemma}
 Lemma \ref{lem1} is a consequence of the following general lemma which will be also useful in the study of the hedge (Section \ref{sec:hedge}).
\begin{lemma}  Let $(\Pi_{\sigma,\tau})_{\sigma \leq \tau}$ be a  No Free Lunch TCPP. Let $\tau$ be a stopping time.
Assume that for some $X \in L^{\infty}(\Omega,{\cal F}_{\tau},P)$ there is $\nu \leq \tau$ and $A \in {\cal F}_{\nu}$ such that: \\
$-\Pi_{\nu,\tau}(-X)1_A=\Pi_{\nu,\tau}(X)1_A $.\\ 
Then for all $\nu \leq \sigma \leq \tau$,
$-\Pi_{\sigma,\tau}(-X)1_A=\Pi_{\sigma,\tau}(X)1_A$.
\label{lemma3.1}
\end{lemma}
{\bf Proof}: For any $\nu \leq \sigma \leq \tau$,
\begin{equation}
-\Pi_{\sigma,\tau}(-X)1_A \leq \Pi_{\sigma,\tau}(X)1_A
\label{eq01}
\end{equation}
 As the TCPP has No Free Lunch, there is a probability measure $R \sim P$ with zero minimal penalty.
From  equations (\ref{eq_NFL2}) and (\ref{eq01})  and time consistency  it follows that  
\begin{eqnarray}
-\Pi_{\nu ,\tau}(-X)1_A \leq E_R(-\Pi_{\sigma,\tau}(-X)1_A|{\cal F}_{\nu})\nonumber\\ 
\leq E_R(\Pi_{\sigma,\tau}(X)1_A|{\cal F}_{\nu}) \leq \Pi_{\nu,\tau}(X)1_A
\label{eq02}
\end{eqnarray}
 By hypothesis $-\Pi_{\nu ,\tau}(-X)1_A=\Pi_{\nu,\tau}(X)1_A$. Thus any inequality in expression (\ref{eq02}) is in fact an equality. As $R \sim P$, it thus follows from (\ref{eq01}) that 
$$-\Pi_{\sigma,\tau}(-X)1_A  = \Pi_{\sigma,\tau}(X)1_A$$ \hfill $\square $
\paragraph{{\bf Proof of Lemma \ref{lem1}}.}
 The equality $C_{bid}(S^k)=C_{ask}(S^k)$ and the hypotheses  1.{\it i} and {\it iii)} of Definition \ref{definition3}
implie that $-\Pi_{0,\tau}(-S^k_{\tau})  = \Pi_{0,\tau}(S^k_{\tau})=S^k_0$. We apply Lemma \ref{lemma3.1} with 
$X= S^k_{\tau}$  $A=\Omega$ and $\nu=0$. It follows that $ -\Pi_{\sigma,\tau}(- S^k_{\tau})= \Pi_{\sigma,\tau}( S^k_{\tau})$. From condition  1.{\it i}, it is also equal to $S^k_{\sigma}$. Thus ii) is proved.
\hfill $\square$
\subsection{Characterization of the calibration}
The following  theorem characterizes the calibration and weak calibration conditions for a No Free Lunch TCPP. 
\begin{theorem}
1. A No Free Lunch TCPP  $(\Pi_{\sigma, \tau})_{0 \leq \sigma \leq \tau}$ is weakly calibrated on  the reference 
family $((S^k)_{0 \leq k \leq d},(Y^l)_{1 \leq l \leq p})$ and the observed limit order books  $(C_{bid}(m S^k)$, $C_{ask}(nS^k))_{1 \leq k \leq d}$, $(C_{bid}(m Y^l),C_{ask}(nY^l)_{1 \leq l \leq p})$ if and only if:\\
-  Local martingale property:\\
Any probability measure $R$ equivalent with $P$ with zero minimal penalty  (i.e. $R \in  {\cal M}^0$)    is an equivalent local martingale measure for the  process  $(S^k)_{0 \leq k \leq d}$.\\
-  Threshold condition: for any  $R \sim  P$,
\begin{eqnarray}
\alpha^m_{0,\tau}(R) & \geq  & \sup_{ \tau_l \leq \tau } (\sup_{m \leq M^l}(C_{bid}(mY^l)-E_R(mY^l)),\sup_{n \leq N^l}((E_R(nY^l)-C_{ask}(nY^l)))\;\;\label{eq_00}\\
\alpha^m_{0,\tau}(R) & \geq  & \sup_{1 \leq k \leq d} |S^k_{0}-E_R(S^k_{\tau})|
\label{eq_Y1}\\
\alpha^m_{0,\tau}(R) & \geq  & \sup_{m \leq M(k)}(C_{bid}(mS^k)-E_R(mS^k_{\tau})),\sup_{ n \leq N(k)}(E_R(nS^k_{\tau})-C_{ask}(nS^k))\;\;\label{eq_Y2}\\ \nonumber
\end{eqnarray}
 2.  A No Free Lunch TCPP  $(\Pi_{\sigma, \tau})_{0 \leq \sigma \leq \tau}$ is calibrated on  the reference 
family $((S^k)_{0 \leq k \leq d}$, $(Y^l)_{1 \leq l \leq p})$ and the limit order books  $(C_{bid}(m Y^l),C_{ask}(nY^l))_{1 \leq l \leq p}$ if and only if  any probability measure $R \in {\cal M}^{1,e}(P)$ (i.e. $R \sim P$ of finite penalty) is a local martingale measure for the process $(S^k)_{0 \leq k \leq p}$, and the threshold condition (\ref{eq_00}) is satisfied. 
\label{thm2.3}   
\end{theorem}
 \begin{remark}:
The fundamental difference between the calibration and the weak calibration for a No Free Lunch TCPP in terms of their dual representation, is the following:\\
 in case of calibration, any probability measure in the dual representation is a local martingale measure for the process $(S^k_t)_{1 \leq k \leq d}$ while in case of weak calibration,  this is only the case for the probability measures with zero penalty.
\label{rq3}
\end{remark}
{\bf Proof of Theorem \ref{thm2.3}}\\
Proof of {\it 1.}\\
-  Assume first that the No Free Lunch TCPP  is weakly calibrated on the reference family. Let $\tau$ be a  stopping time such that the stopped process $(S^k)^{\tau}_{0 \leq k \leq d}$ is uniformly bounded. Let $0 \leq \sigma \leq \tau$.  $\Pi_{\sigma,\tau}(S^k_{\tau})=-\Pi_{\sigma,\tau}(-S^k_{\tau})=S^k_{\sigma}$. Let $R \in {\cal M}^0$,  $\alpha^m_{\sigma,\tau}(R)=0$. From the dual representation, equation (\ref{eq_NFL2}), it follows that 
$$ S^k_{\sigma}=-\Pi_{\sigma,\tau}(-S^k_{\tau}) \leq E_R(S^k_{\tau}|{\cal F}_{\sigma})\leq \Pi_{\sigma,\tau}(S^k_{\tau})=S^k_{\sigma}\;$$
Thus any $R$ equivalent with $P$ with zero minimal penalty is a local martingale measure for $(S^k)_{0 \leq k \leq d}$.
The threshold condition follows from the expression of the minimal penalty $\alpha^m_{0,\tau}(Q)=\sup_{Z \in L^{\infty}({\cal F}_{\tau})}(E_Q(Z)-\Pi_{0,\tau}(Z))$
\\
-  Conversely, assume that the No Free Lunch TCPP satisfies the local martingale property and the threshold condition. We have to prove  that the pricing procedure satisfies the conditions   of definition \ref{definition3}.\\
 Let $\tau$ be a  stopping time such that the stopped process $(S^k)^{\tau}_{0 \leq k \leq d}$ is uniformly bounded. Let $R \in {\cal M}^0$. From the dual representation of $\Pi_{\sigma,\tau}$, equation (\ref{eq_NFL2}), as $R$ is a local martingale measure for $(S^k)_{0 \leq k \leq d}$, it follows then that  
\begin{equation}
\forall \sigma \leq \tau\;\;\forall n \in \N\;\;-\Pi_{\sigma,\tau}(-n S^k_{\tau}) \leq  nS^k_{\sigma} \leq \Pi_{\sigma,\tau}(n S^k_{\tau})
\label{eqE}
\end{equation}
Thus condition 1. {\it i)} of Definition \ref{definition3} is satisfied.
 From the threshold condition, for $n \leq N(k)$, for any  $Q \in {\cal M}^{1,e}(P)$, $E_Q(nS^k_{\tau})-\alpha^m_{0,\tau}(Q)\leq C_{ask}(nS^k)$. So applying the equation of representation (\ref{eq_NFL2}) to 
 $\Pi_{0,\tau}$, we get $$ \Pi_{0,\tau}(n S^k_{\tau}) \leq C_{ask}(nS^k)\;\;\forall n \leq N(k)$$
 The inequality $ C_{bid}(mS^k) \leq  -\Pi_{0,\tau}(-m S^k_{\tau})\;\forall m \leq M(k)$  is proved in the same way. Thus conditions 1.{\it iii)} of Definition \ref{definition3} is satisfied.\\
The proof of condition  2.  of Definition \ref{definition3} is analogous.\\
We prove now that the condition 1. {\it ii)} is satisfied. 
From equation (\ref{eqE}), $ -\Pi_{0,\tau}(-S^k_{\tau}) \leq S^k_{0} \leq \Pi_{0,\tau}(S^k_{\tau})$.  The converse inequality is a consequence of the threshold condition (inequation (\ref{eq_Y1})) and of the equation of representation (\ref{eq_NFL2}) applied to $\Pi_{0,\tau}$. Thus  $S^k_0=\Pi_{0,\tau}(S^k_{\tau})= -\Pi_{0,\tau}(-S^k_{\tau})$. Applying now Lemma \ref{lemma3.1}   , we get the equality 
 $$\Pi_{\sigma,\tau}(S^k_{\tau})= -\Pi_{\sigma,\tau}(-S^k_{\tau})$$
Using  (\ref{eqE}) it is also equal to $S^k_{\sigma}$ thus  condition 1. {\it ii)}  of Definition \ref{definition3} is satified. This proves {\it 1} .\\
For the proof of {\it 2} we refer to \cite{BN05}. \hfill $\square $
\subsection{Extended Version of Kreps Yan  First Fundamental Theorem}
\label{sec:EV} 
 We state now an extended version of  Theorem \ref{thm1} of Section \ref{sec:fund. Thm}.
\begin{theorem}
The following conditions are equivalent:
\begin{itemize}
\item[i)] The reference family  $((S^k)_{0 \leq k \leq d},(Y^l)_{1 \leq l \leq p})$ satisfies the No Free Lunch condition with respect to the limit order books  $(C_{bid}(mY^l), C_{ask}(nY^l))$.
\item[ii)]
 There is an equivalent local martingale measure $R$ for $(S^k)_{0 \leq k \leq d}$ such that for any 
$ l \in \{1,...,p\}$, for $m \leq M^l\;\;C_{bid}(mY^l) \leq E_R(mY^l)$ and for $n \leq N^l\;\;E_R(nY^l) \leq C_{ask}(nY^l)$.
\item[iii)] There is a No Free Lunch TCPP calibrated on the reference family $(S^k_{0 \leq k \leq d}$, $Y^l_{1 \leq l \leq p})$  and the limit order books  $(C_{bid}(mY^l), C_{ask}(nY^l))$.
\item[iv)]  There is a No Free Lunch TCPP weakly calibrated on the reference \\family $(S^k_{0 \leq k \leq d},Y^l_{1 \leq l \leq p})$  and the limit order books
 $(C_{bid}(mS^k), C_{ask}(nS^k))$, $(C_{bid}(mY^l),$ $C_{ask}(nY^l))$.
\end{itemize}
\label{thm2}
\end{theorem}
The proof of this extended version of the First Fundamental Theorem is given in Appendix \ref{secEFFT}.  A key tool in this proof is  the existence of an equivalent probability measure with zero minimal penalty. The aim of the proof is the same as that of the proof of  Kreps Yan Theorem given in \cite{DS02}.\\
Theorem \ref{thm2} shows also that the notion of No Free Lunch TCPP  is well adapted to the questions related to No Arbitrage. 
\section{Properties of the Supply Curve}
\label{sec:supply}
 Let $(\Pi_{\sigma,\tau})_{\sigma \leq \tau}$ be a No Free Lunch TCPP. Let $X$ be an essentially bounded non negative financial asset.
For $x \in \R^+*$,(resp  $x \in \R^-*$) denote $X(t,x,\omega)$ the ask price (resp bid price) at time $t$ per share for an order of size $x$, which means that $X(t,x,\omega)=\frac{\Pi_{t,\infty}(xX)(\omega)}{x}$.
In the following proposition we list the properties of the supply  curve.
\begin{proposition}
\begin{enumerate}
\item For any $x$, $(t,\omega) \rightarrow X(t,x,\omega)$ is a c\`adl\`ag stochastic process.
\item There is an equivalent probability measure $R$ such that for any $x \geq 0$, the process $X(t,x,.)$ is a $R$-supermartingale and for any $x\leq 0$ the process  $X(t,x,.)$ is a $R$-submartingale.
\item For any $\tau$, $x\in \R^* \rightarrow X(\tau,x,.) \in L^{\infty}(\Omega,{\cal F}_{\tau},P)$ is non decreasing.
$\forall \tau $, $P\;a.s.$, $x \rightarrow X(\tau,x,\omega)$ is  continuous, admits a right and a left derivative at any point. It is twice derivable almost surely. 
\item limit in $zero$:
For any $\tau$,  $x \rightarrow X(\tau,x,.)$ has a right (resp. a left) limit in $0$ in $L^{\infty}(\Omega,{\cal F}_{\tau},P)$ denoted $X^+(\tau,0,.)$ (resp.$X^-(\tau,0,.)$ 
\begin{eqnarray}
X^+(\tau,0,.)=\esssup_{Q \in {\cal M}^0}(E_Q(X|{\cal F}_{\tau})\nonumber\\
X^-(\tau,0,.)=\essinf_{Q \in {\cal M}^0}(E_Q(X|{\cal F}_{\tau})
\label{eqzero}
\end{eqnarray}
where ${\cal M}^0$ is the set of probability measures with zero minimal penalty in the dual representation of the TCPP (equations (\ref{eq_NFL2}) and (\ref{eqzeropen})).
\item Asymptotic limit:
For any $\tau \in \R^+$, $X(\tau,x,.)$ has a limit as  $x \rightarrow +\infty$ (resp $x \rightarrow -\infty$) denoted $X^{\infty}(\tau,.)$ (resp $X^{-\infty}(\tau,.)$). $X^{\infty}(\tau,.)$  and $X^{-\infty}(\tau,.)$ are   c\`adl\`ag processes.  
\begin{eqnarray}
X^{\infty}(\tau,.)=\esssup_{Q \in  {\cal M}^{1,e}(P)}(E_Q(X|{\cal F}_{\tau})\\
X^{-\infty}(\tau,.)=\essinf_{Q \in  {\cal M}^{1,e}(P)}(E_Q(X|{\cal F}_{\tau})
\label{eqas}
\end{eqnarray}
with the notations of (\ref{eq_NFL2}) and (\ref{eq_2NFL}).
\end{enumerate}
\label{propsup}
\end{proposition}
{\bf Proof}.
 As the TCPP has No Free Lunch, it follows from \cite{BN05} that ${\cal M}_0$ is non empty. let $R \in {\cal M}_0$, {\it 1} and {\it 2} follow then from \cite{BN04} Corollary 1 of  Theorem 3.\\
{\it 3}. follows from the convexity of $\Pi_{\tau,\infty}$ and normalization (i.e.
$\Pi_{\tau,\infty}(0)=0)$\\
{\it 4}. Let  $X \in L^{\infty}(\Omega,{\cal F}_{\infty},P)$,
$\forall x \in \R^+*$, $\frac{\Pi_{\tau,\infty}(xX)}{x} \geq \esssup_{Q \in {\cal M}^0}(E_Q(X|{\cal F}_{\tau})$, so 
\begin{equation} X^+(\tau,0,.) \geq \esssup_{Q \in {\cal M}^0}(E_Q(X|{\cal F}_{\tau})
\label{eq0001}
\end{equation}
\begin{equation}
E_P( X^+(\tau,0,.) \leq \inf_{x \in \R^+*}(\sup_{Q \in {\cal M}^{1,e}(P)}(E_P(E_Q(X|{\cal F}_{\tau})-\frac{E_P(\alpha^m_{\tau,\infty}(Q)}{x}))
\label{eq00_0}
\end{equation}
If $E_P(\alpha^m_{\tau,\infty}(Q)) \neq 0$, $\frac{E_P(\alpha^m_{\tau,\infty}(Q))}{x} \rightarrow \infty$ as $x \rightarrow 0$. Therefore we can restrict in (\ref{eq00_0}) to probability measures $Q\sim P$ such that $\alpha^m_{\tau,\infty}(Q)=0\;P\;a.s.$. Choose $R \in {\cal M}^0$. denote $\tilde Q$ the probability measure of Radon Nykodim derivative
$$\frac{d \tilde Q}{dP}=E(\frac{dR}{dP}|{\cal F}_{\sigma})
\frac{\frac{dQ}{dP}}{E(\frac{dQ}{dP}|{\cal F}_{\sigma})}$$
for all $ X$, $ E_Q(X|{\cal F}_{\tau})=  E_{\tilde Q}(X|{\cal F}_{\tau})$ and $\tilde Q \in {\cal M}^0$.
Thus\\ $E_P( X^+(\tau,0,.)) \leq \inf_{x \in \R^+*}(\sup_{\tilde Q \in {\cal M}^0}E_{\tilde Q}(X))$. {\it 4.} follows then from (\ref{eq0001}).\\
{\it 5}. The increasing limit of $X(t,x)$ as $x \rightarrow \infty$, defines a sublinear  No Free Lunch TCPP. Denote it $\Pi^{\infty}_{\sigma,\tau}$. From the dual representation of $\Pi$, and the non negativity of the minimal penalty, it follows that for any $X \geq 0$, for any $x \in \R^+$, $\frac{\Pi_{\sigma,\tau}(xX)}{x} \leq \esssup_{Q \in  {\cal M}^{1,e}(P)}(E_Q(X|{\cal F}_{\sigma})$. Thus 
 $(\Pi^{\infty})_{\sigma,\tau}(X) \leq \esssup_{Q \in  {\cal M}^{1,e}(P)}(E_Q(X|{\cal F}_{\sigma})$.
For any $Q \in {\cal M}^{1,e}(P)$, $\alpha_{0,\infty}(Q) < \infty$ 
$\alpha^{\infty}_{0,\infty}(Q)=\sup_{Y \in L^{\infty}(\Omega,{\cal F}_{\infty},P)}(E_Q(Y)-\Pi^{\infty}_{0,\infty}(Y))$ From the inequality $\Pi_{0,\infty}(Y)) \leq \Pi^{\infty}_{0,\infty}(Y))$, it follows that $\alpha^{\infty}_{0,\infty}(Q) < \infty$. $\Pi^{\infty}$ is sublinear so $\alpha^{\infty}_{0,\infty}(Q)=0$. Thus   $(\Pi^{\infty})_{\sigma,\tau}(X) \geq \esssup_{Q \in  {\cal M}^{1,e}(P)}(E_Q(X|{\cal F}_{\sigma})$ and {\it 5.} is proved. \hfill $\square $

In the particular case where the TCPP is calibrated  on the reference family we get the following corollary. 
\begin{corollary}
Assume that the TCPP is calibrated on the reference family  $((S^k)_{0 \leq k \leq d},$ $(Y^l)_{1 \leq l \leq p})$  and the limit order book $(C_{bid}(mY^l),C_{ask}(nY^l))_{1 \leq l \leq p}$. 
\begin{enumerate}
\item For any $k$, $S^k(t,x,\omega)=S^k(t,\omega)$
\item  Assume that  there is a transaction at time $0$ on the option $Y^l$, then the process $Y^l(t,x,\omega)$ has a limit as $x$ tends to $0$, $\forall t, \;\; (Y^l)^+(t,0,.)=(Y^l)^-(t,0,.)=\Pi_{t,\tau_l}(Y^l)$
\item For any financial instrument $X$, the asymptotic limit  $X^{\infty}(t,.)$is less or equal to the surreplication price (with respect to the basic assets  $((S^k)_{0 \leq k \leq d})$, i.e. $$X^{\infty}(t,.)\leq \esssup_{Q \in {\cal M}(S)}E_Q(X|{\cal F}_t)(\omega)$$ where ${\cal M}(S)$ denotes the set of all equivalent local martingale measures for the process $S=(S^k)$.
 There is equality in the above equation if and only if the set of probability measures ${\cal M}^{1,e}(P)$ in the dual representation of the TCPP is equal to ${\cal M}(S)$ (with the notations of (\ref{eq_NFL2}) and (\ref{eq_2NFL}).
\end{enumerate} 
\end{corollary}

{\bf Proof}.
{\it 1} follows from definition of calibration.\\
{\it 2} As there is a transaction at time $0$ on $Y^l$, $C_{bid}(Y^l)=c_{ask}(Y^l)=\Pi_{0,\infty}(Y^l)=-\Pi_{0,\infty}(-Y^l)$. Thus,from Lemma \ref{lemma3.1}, for any $t\geq 0$, $\Pi_{t,\infty}(Y^l)=-\Pi_{t,\infty}(-Y^l)$
 $\forall x \in ]0,1[,\;\;-\Pi_{t,\infty}(-Y^l)\leq \frac{\Pi_{t,\infty}(-xY^l)}{-x} \leq \frac{\Pi_{t,\infty}(xY^l)}{x} \leq \Pi_{t,\infty}(Y^l)$
Therefore $(Y^l)^+(t,0,\omega)=\Pi_{t,\infty}(Y^l)=\Pi_{t,\tau_l}(Y^l)$\\
{\it 3} follows from Theorem \ref{thm2.3} and Proposition \ref{propsup}.
\begin{remark}In this paper we work in a very general framework of illiquid market represented by a general filtered probability space. From a very simple axiomatic for TCPP we have proved properties (Proposition (\ref{propsup})) satisfied by the supply curve associated with any financial product. We can compare these  properties with the properties which were assumed in \cite{CJP} for one asset. Notice first that our model is an infinite dimensional model. The supply curve is defined for any asset i.e; any essentially bounded random variable. We list the main differences for the supply curve:\\ 
-We have proved that the sample paths are c\`adl\`ag (and therefore allow for jumps) whereas in \cite{CJP} the sample paths were assumed to be continuous.\\
 We do not have in general a limit as $x$ tends to $0$ for $S(t,x,.)$ but only a right limit and a left limit. \\
A strong hypothesis of smoothness is made in \cite{CJP}: $S(t,x, \omega)$ is assumed to be of $C^2$ class in $x$.  We get here that it is  twice derivable almost surely in $x$.
\end{remark}

\section{Robustness for TCPP calibrated on option prices}
\label{sec:stable}
In this section we study the robustness  of TCPP calibrated on the reference family $((S^k)_{1 \leq k \leq d},(Y^l)_{1 \leq l \leq d})$ and the limit order books $((C_{bid}(mY^l))_{m \leq M^l}$, $(C_{ask}(nY^l))_{n \leq N^l}$. 
\begin{proposition}The maximal bid-ask interval associated to No Free Lunch TCPP calibrated on the reference family $((S^k)_{1 \leq k \leq d},(Y^l)_{1 \leq l \leq d})$ and the limit order books $((C_{bid}(mY^l))_{m \leq M^l},(C_{ask}(nY^l))_{n \leq N^l}$ is given, for any financial asset $X$, by
$$[m_X,M_X]=[\inf_{Q \in {\cal M}_e}(E_Q(X)+\beta(Q),\sup_{Q \in {\cal M}_e}(E_Q(X)-\beta(Q)]$$
with 
\begin{equation}\beta(Q)=\sup_{l}[\sup_{m \leq M^l}( (C_{bid}(mY^l)-E_Q(mY^l),\sup_{n \leq N^l}E_Q(nY^l))-C_{ask}(nY^l)]
\label{beta}
\end{equation}
where ${\cal M}_e$ is the set of equivalent local martingale measures for $(S^k)_{1 \leq k \leq d}$
\label{prop:stable}
\end{proposition}
{\bf Proof}. This results from Theorem \ref{thm2.3}.\\
It follows from this Proposition that a little move for  $C_{bid}(nY^l)$ and $C_{ask}(nY^l)$ induces for any $X$ a small change in the maximal bid-ask spread associated with $X$.
Indeed, denote $\beta'(Q)$ the minimal penalty associated to No Free lunch TCPP calibrated on the limit order books  $((C'_{bid}(mY^l))_{m \leq M^l}$, $(C'_{ask}(nY^l))_{n \leq N^l}$.  let $\epsilon$ such that  $\epsilon \geq |(C_{bid}(mY^l)-(C_{bid}(mY^l)|$ for $m\leq M^l$ and $\epsilon \geq |(C_{ask}(nY^l)-(C'_{ask}(nY^l)|$ for  $n \leq N^l$.  From equation (\ref{beta}), it follows that $|\beta(Q)-\beta'(Q)|\leq \epsilon$ and thus $|m_X-m'_X| \leq \epsilon$, $|M_X-M'_X| \leq \epsilon$ for any $X$.

\section{A hedging result for TCPP calibrated on liquid options}
\label{sec:hedge}
The aim of this Section is to study  No Free Lunch TCPP calibrated on perfectly liquid options $Y^l$,  and to prove a hedging result.\\
We assume that for any of the reference options $(Y^l)_{1 \leq l \leq d}$, $C_{bid}(nY^l)=C_{ask}(nY^l)$. Denote  $C^l=C_{bid}(Y^l)=C_{ask}(Y^l)$. From the convexity of $n \rightarrow C_{ask}(nY^l)$, (and concavity of $n \rightarrow C_{bid}(nY^l)$ it follows that $n C^l=C_{ask}(nY^l)=C_{bid}(nY^l)$. In that case we simply say that the TCPP is   calibrated on   the 
reference family $((S^k)_{0 \leq k \leq d},(Y^l)_{1 \leq l \leq p})$ and the observed  prices $({C^l})_{1 \leq l \leq p}$. Let $(\Pi_{\sigma,\tau})_{\sigma \leq \tau}$ be such a No Free Lunch TCPP.
From equation (\ref{eq_NFL2}) and Theorem \ref{thm2.3} , it follows that there is a  set ${\cal Q}$ of equivalent local martingale  measures for $(S^k)_{0 \leq k \leq d}$, with $\alpha_{0,\infty}(Q)<\infty$,  such  that 
\begin{equation}
\forall \sigma \leq \tau\;\;\Pi_{\sigma,\tau}(X)=\esssup_{Q \in {\cal Q}}(E_Q(X|{\cal F}_{\sigma})-\alpha^m_{\sigma,\tau}(Q))
\label{eqPi}
\end{equation}
We say that the No Free Lunch TCPP is  represented by the set ${\cal Q}$.\\The following lemma is a key  result for the study of the hedge:  the process $Z^l_t=\Pi_{t,\infty}(Y^l)$ is a martingale for any  $Q$ in ${\cal Q}$.  
\begin{lemma}
Let  $(\Pi_{\sigma,\tau})_{\sigma \leq \tau}$ be a  No Free Lunch TCPP calibrated on the reference family  $((S^k)_{0 \leq k \leq d},(Y^l)_{1 \leq l \leq p})$  and the  prices $(C^l)_{1 \leq l \leq p}$. Then
$$\forall 1 \leq l \leq p,\;\;\forall \sigma,\;\; \forall n \in \N, \;\;\Pi_{\sigma,\infty}(nY^l)=-\Pi_{\sigma,\infty}(-nY^l)$$
 The process $Z^l_{t}=\Pi_{t,\infty}(Y^l)$ is a martingale with respect to any probability measure in ${\cal Q}$.
\label{lemmahedge}
\end{lemma}
{\bf Proof}. For all  $n \in \N$, $-\Pi_{0,\infty}(-nY^l)=\Pi_{0,\infty}(nY^l)=nC^l$. From Lemma \ref{lemma3.1} applied with $\nu=0$, it follows that  for any  $\sigma$,
\begin{equation}
-\Pi_{\sigma,\infty}(-nY^l)=\Pi_{\sigma,\infty}(nY^l) 
\label {eqn}
\end{equation}
The convexity of $\Pi_{\sigma,\infty}$ implies  that $$-\Pi_{\sigma,\infty}(-nY^l) \leq  -n \Pi_{\sigma,\infty}(-Y^l)\leq  n \Pi_{\sigma,\infty}(Y^l) \leq \Pi_{\sigma,\infty}(nY^l)$$
 From (\ref{eqn}), it follows that any inequality in the above relation is in fact an equality. Thus $ n \Pi_{\sigma,\infty}(Y^l)=\Pi_{\sigma,\infty}(nY^l) \;\forall n \in \Z$. From equation (\ref{eqPi}), it  follows that $\forall Q \in {\cal Q},\;\;\forall n \in \N$, $\alpha^m_{\sigma, \infty}(Q)\geq n |\Pi_{\sigma,\infty}(Y^l)-E_Q(Y^l|{\cal F}_{\sigma})|$ a.s. From the cocycle equation (\ref{eqcoc}), $\forall Q \in {\cal Q}\;\; E_Q(\alpha^m_{\sigma,\infty}(Q))<\infty$.\\
Then $Z^l_{\sigma}= \Pi_{\sigma,\infty}(Y^l)=E_Q(Y^l|{\cal F}_{\sigma})= E_Q(Z^l_{\tau}|{\cal F}_{\sigma}) a.s.,\;\;\forall \tau \geq \sigma$.
\hfill $\square $
\begin{theorem} Let  $(\Pi_{\sigma,\tau})_{\sigma \leq \tau}$ be a  No Free Lunch TCPP calibrated on the reference family  $((S^k)_{0 \leq k \leq d},(Y^l)_{1 \leq l \leq p})$  and the observed prices $(C^l)_{1 \leq l \leq p}$. Then $\forall X \in L^{\infty}(\Omega, {\cal F}, P)$, 
\begin{equation}
\Pi_{0,\infty}(X)  \leq  \inf \{x |\;there \;is\; h \in {\cal C}\; with \;x+h = X \}
\label{eqhedge}
\end{equation}
where ${\cal C}=(K_0-L^0_+)\INTER L^{\infty}$ and
\begin{eqnarray}
K_0=\{(H.S)_{\infty}+(K.Z)_{\infty}\;|\;H, K \;admissible \;and \; \nonumber \\(H.S)_{\infty}=\lim_{t \rightarrow \infty}(H.S)_{t}\;exists \;a.s. \;idem\; for\; K.Z\}\nonumber
\end{eqnarray}
This gives a better superhedge result than the usual one.
\label{thm3}
\end{theorem}
{\bf Proof}.
 Denote  ${\cal M}^e(S^k,Z^l)$ the set of equivalent local martingale measures for the process $(S^k,Y^l)$. From Lemma \ref{lemmahedge},  and part {\it 2.} of Theorem \ref{thm2.3}, any probability measure in ${\cal Q}$ is an equivalent local martingale measure for $(S^k,Z^l)$, i.e. ${\cal Q} \subset {\cal M}^e(S^k,Z^l)$.  Therefore from equality (\ref{eqPi}), we get 
$\Pi_{0,\infty}(X) \leq \sup_{Q \in {\cal M}^e(S^k,Z^l)}E_Q(X)$. One can now apply the superhedge result  of Delbaen and Schachermayer \cite{DS01} 
( Theorem 9.5.8 in \cite{DS02}). This proves (\ref{eqhedge}).
\hfill $\square $\\ 
Economic interpretation of this result:
  When the options $Y^l$ are perfectly liquid, a TCPP $(\Pi)_{\sigma,\tau}$
 calibrated on the reference family  $((S^k)_{0 \leq k \leq d},(Y^l)_{1 \leq l \leq p})$  and the  prices $(C^l)_{1 \leq l \leq p}$ is a TCPP extending the dynamics of the processes $((S^k_t)_{0 \leq k \leq p}, Z^l_t=E_Q(Y^l|{\cal F}_t)_{1 \leq l \leq d})$ where $Q$ is any equivalent probability measure involved in the representation of $(\Pi)_{\sigma,\tau}$ (equation (\ref{eqPi})).  The options $Y^l$ can be used to hedge dynamically as well as the assets $S^k$.
\begin{definition}
The TCPP $\Pi^0$ is said maximal among the TCPP calibrated on the reference family $(S^k,Y^l)$ and the prices $(C^l)$ if for any TCPP $\Pi$ calibrated on  $(S^k,Y^l)$ and the prices $(C^l)$, for any $Y \geq 0$, $\Pi(Y) \leq \Pi^0(Y)$. 
\end{definition}
From Theorem \ref{thm3}, we deduce the following result:
\begin{corollary}
Let $\Pi^0$ be a maximal TCPP calibrated on  $(S^k,Y^l)$ and the prices $(C^l)$ then $\Pi^0$ is sublinear and represented by the set  of all equivalent local martingal measures for the process $((S^k,Z^l))$  ${\cal M}^e(S^k,Z^l)$. The inequality (\ref{eqhedge}) becomes  an equality, leading to a perfect hedge result. 
\end{corollary}

\section{TCPP in a stochastic volatility model}

In this Section we assume that the price process $S$ of a primitive asset expressed in terms of the num\'eraire $S^0$ satisfies a stochastic volatility model.  We assume that for any of the $Y^l$ one observes a limit order book. The notations for the limit order book are those of Subsection \ref{sec:book}.\\
Assuming that the reference family satisfies the No Free Lunch condition, there is (Theorem \ref{thm1}) an equivalent local martingale measure $R$ for the process $S$ such that for any $l$, $C^l_{bid} \leq E_R(Y^l) \leq C^l_{ask}$. We take this probability measure $R$ as the new reference probability.
Therefore we assume that the price process $S$ of the primitive asset expressed in terms of the num\'eraire $S^0$ is given by
\[
\left \{ \parbox{4cm}{
\begin{eqnarray} 
\frac {dS_t}{S_t} & = & \sigma_t(\sqrt{1-\rho^2}dW^1_t+\rho dW^2_t) \nonumber\\
d\sigma_t & = & \alpha(t,S_t,\sigma_t)dt+\gamma(t,S_t,\sigma_t)dW^2_t\nonumber
\end{eqnarray}
}\right.
\]
where $W^1$ and $W^2$ are two independent Brownian motions and $\rho \in ]-1,1[$.
  \begin{proposition}
Any TCPP calibrated on the reference family $(S,(Y^l)_{1 \leq l \leq p})$ can be written
$$ \Pi_{\sigma,\tau}(X)=\esssup_{\{\nu | \int_{0}^{\infty}\nu_s^2 ds <\infty\}} (E_{Q_\nu}(X|{\cal F}_{\sigma})-\alpha_{\sigma,\tau}(Q_{\nu}))$$

with 
\begin{equation}
\frac{dQ_{\nu}}{dR}=exp(-\int_{0}^{\infty}\frac{\rho \nu_s}{\sqrt{1-\rho^2}} dW^1_s+\int_{0}^{\infty}\nu_s dW^2_s-\frac{1}{2}\int_{0}^{\infty}\frac{ \nu_s^2}{{1-\rho^2}} ds)
\label{Eqsv}
\end{equation}
Furthermore for any such TCPP for any $X \in L^{\infty}(\Omega,{\cal F}_t,P)$, 
$$ \Pi_{0,t}(X) \leq \sup_{\nu} (E_{Q_\nu}(X)-\alpha^m_{0,t}(Q_{\nu}))$$
with 
\begin{equation}
\alpha^m_{0,\tau}(Q_{\nu})= \sup(0,\sup_{ \tau_l \leq \tau } (\sup_{n \leq M^l}C_{bid}(nY^l)-E_{Q_{\nu}}(nY^l),\sup_{n \leq N^l}E_{Q_{\nu}}(nY^l)-C_{ask}(nY^l))\label{eq_l}\\
\end{equation}
\label{prop7.1.4}
\end{proposition}
{\bf Proof}.
From Theorem \ref{thm2.3}, any TCPP extending the dynamics of $S$ has a representation in terms of equivalent local martingale measures for $S$.
Any probability measure equivalent with $R$ is characterized by its Radon Nikodym derivative
\begin{equation}
\frac{dQ}{dR}=exp(\int_{0}^{\infty}\lambda_s dW^1_s-\frac{1}{2}\int_{0}^{\infty}\lambda_s^2 ds+\int_{0}^{\infty}\nu_s dW^2_s-\frac{1}{2}\int_{0}^{\infty}\nu_s^2 ds)
\label{eqsto0}
\end{equation}
Let $W^1_t(Q)=W^1_t - \int_{0}^{\infty}\lambda_s ds$ and $W^2_t(Q)=W^2_t - \int_{0}^{\infty}\nu_s ds$.
 From Girsanov's Theorem, $(W^1_t(Q),W^2_t(Q))$ is a two dimensional Brownian motion under the probability measure $Q$ and the dynamics of $S$ can be written
\[
\left \{\parbox{13cm}{
\begin{eqnarray} 
\frac {dS_t}{S_t} & = &  \sigma_t[(\lambda_t\sqrt{1-\rho^2}+\rho \nu_t)dt +(\sqrt{1-\rho^2}dW^1_t(Q)+\rho dW^2_t(Q))] \nonumber\\
d\sigma_t & = & (\alpha(t,S_t,\sigma_t)+ \nu_t \gamma(t,S_t,\sigma_t))dt + \gamma(t,S_t,\sigma_t)) dW^2_t(Q) \nonumber
\end{eqnarray}
}\right.
\]
Therefore $Q$ is a local martingale measure for $S$ if and only if $\lambda_t\sqrt{1-\rho^2}+\rho \nu_t=0\;\; a.s.$. Let $Q_{\nu}$ be  the equivalent local martingale measure of Radon Nikodym derivative given by the formula (\ref{eqsto0}) with $ \lambda_t=-\frac{\rho \nu_t}{\sqrt{1-\rho^2}}$. Then $Q_{\nu}$ satisfies equation (\ref{Eqsv}).
From Theorem \ref{thm2.3}, the minimal penalty has to satisfy the threshold condition (\ref{eq_00}), leading then to the result.

 \section{TCPP calibrated on options from BMO martingales}
\label{sec:BMO} 

\subsection{General construction of TCPP calibrated on a reference family}
\label{sec:TCPP} 
In Section \ref{sec:admissibility}, we have characterized  No Free Lunch TCPP calibrated on a reference family in terms of their dual representation. The next step is to construct such No Free Lunch TCPP in a very general setting where the processes $(S^k)_{0 \leq k \leq d}$ can allow for jumps.
 Assume that the reference family satisfies the No Free Lunch condition.
Denote ${\cal M}$ the set of equivalent local martingale measure  for $(S^k)_{0 \leq k \leq d}$ and
$${\cal M}_1= \{R \in {\cal M} \;|\forall l \in \{1,...,p\}\;\;C_{bid}(nY^l) \leq E_R(nY^l) \leq C_{ask}(nY^l)\}$$
 Let $R \in {\cal M}_1$. The conditional expectation with respect to $R$ provides thus a linear TCPP calibrated on the reference family. However as soon as one calibrates on options which are not perfectly liquid, one has to construct non linear, and even more, non sublinear TCPP in order to take care of the non liquidity. One has introduced in  ~\cite{BN03} a general method to construct convex TCPP starting with a stable set of equivalent probability measures and defining on it a penalty. In general the penalty constructed is not the minimal one. The following result gives  sufficient conditions on the set of probability measures and on the penalty for the construction of TCPP calibrated on the reference family.  
 For the definition of stability of a set of probability measures we refer to  ~\cite{D} and ~\cite{BN04}.
 For the definition of locality  for the penalty $\alpha$ we refer to  ~\cite{BN04} definition 5. The cocycle condition is given by the equation (\ref{eqcoc})
with $\alpha$ instead of $\alpha^m$.
\begin{proposition}
Let ${\cal M}$ be a stable set of probability measures all equivalent to $P$.
Let $\alpha$  be a  non negative penalty function on ${\cal M}$. Assume that there is $Q \in {\cal M}$ such that $\alpha_{0,\infty}(Q)=0$. 
Assume that the penalty function $\alpha$ is  
local 
and satisfies the cocycle condition.
\\Consider the No Free Lunch TCPP $(\Pi_{\sigma,\tau})_{\sigma \leq \tau}$ defined by
\begin{equation}
\forall \; X \in L^{\infty}({\cal F}_{\tau})\;\;\Pi_{\sigma,\tau}(X)=\esssup_{Q \in {\cal M}} (E_Q(X|{\cal F}_{\sigma})-\alpha_{\sigma,\tau}(Q))
\label{eqF}
\end{equation}
\begin{enumerate}
\item  The TCPP is calibrated on the  reference family  $((S^k)_{0 \leq k \leq d},(Y^l)_{1 \leq l \leq p})$  and the limit order books   $(C_{bid}(mY^l))_{m \leq M^l},(C_{ask}(nY^l)_{n \leq N^l})_{1 \leq l \leq p})$ if  it satisfies the two following conditions:\\
 i) Local martingale property: Any element of ${\cal M}$ is an equivalent local martingale measure for $(S^k)_{0 \leq k \leq d}$ \\
  ii) threshold condition for the penalty: for any  $R\in {\cal M}$,
\begin{equation}
\alpha_{0,\tau}(R)  \geq   \sup_{ \tau_l \leq \tau } (\sup_{m \leq M^l}C_{bid}(mY^l)-E_R(mY^l),\sup_{n \leq N^l}E_R(nY^l)-C_{ask}(nY^l))
\label{eq_00bis}
\end{equation}
\item It is weakly calibrated on the reference family if\\
i') any  $R \in {\cal M}$ with zero penalty is an equivalent local martingale measure for $S^k$.\\
ii') threshold condition: for any  $R\in {\cal M}$, inequality (\ref{eq_00bis}) is satisfied as well as (\ref{eq_Y1}) and (\ref{eq_Y2}) with $\alpha_{0,\tau}$ instead of  $\alpha^m_{0,\tau}$. 
\end{enumerate}
\label{propStable1}
\end{proposition}
{\bf Proof}.
 From Theorem 4.4 of~\cite{BN03} and its extended version Proposition 3 of ~\cite{BN04}, formula (\ref{eqF}) defines a time consistent  dynamic pricing procedure. $\alpha_{0,\infty}(Q)=0$, thus the minimal penalty $\alpha^m_{0,\infty}(Q)$ is also equal to $0$. Therefore the TCPP   has No Free Lunch.\\
Notice that the part of the proof of Theorem \ref{thm2.3} starting with ``conversely'' does not use the specific expression of the minimal penalty. It applies to any penalty. This proves {\it 1} and {\it 2}. \hfill $\square $\\
 In order to construct a No Free Lunch TCPP calibrated on the reference family,, we start with the construction of a stable set of equivalent local martingale measures for  $S^k$. This is the easy part. Then we have to prove the existence of penalties satisfying the threshold condition  inequality (\ref{eq_00bis}), this is the difficult part. One has to prove that the bound is satisfied uniformly for any $R \in {\cal M}$. The examples of TCPP that we construct here belong all to the new class that we  first introduced in~\cite{BN03}, using right continuous BMO martingales. For the theory of right continuous BMO martingales we refer to  Dol\'eans-Dade and Meyer \cite{DDM01}.
\subsection{A generic family  of convex TCPP calibrated on option prices}
\label{sec ex-admissible}
Assume that the locally bounded process $(S^k)_{0 \leq k \leq d}$ satisfies the usual No Free Lunch condition. Let $Q_0$ be a local martingale measure for $(S^k)$. For simplicity one assumes that  $(S^k)_{0 \leq k \leq d}$ is a square integrable martingale with respect to $Q_0$. 
\begin{proposition}
 Let $M^1,...,M^j$ be  strongly orthogonal square integrable right continuous martingales in $(\Omega,{\cal F},({\cal F}_t)_{0 \leq t},P)$. Assume that each $M^i$ is furthermore strongly orthogonal to the  martingale  
$(S^k)_{1 \leq k \leq d}$.   Let $(\Phi_i)_{1 \leq i \leq j}$ be a  non negative predictable processes such that the stochastic integral $\Phi. M^i$ is a BMO martingale of BMO norm  $m^i$. Any martingale  in  $${\cal M}=\{\sum_{1 \leq i \leq j} H_i.M^i, \;\;\;H_i\;\;predictable \;\;|H_i| \leq \Phi_i \;\; a.s.\}$$ is   BMO  with  BMO norm bounded by $(\sum _{1 \leq i \leq j} (m^i)^2)^\frac{1}{2}=m$.\\
If  $m<\frac{1}{16}$, ${\cal Q}({\cal M})=\{Q_M\;;\;\frac{dQ_M}{dP}={\cal E}(M)\;|\; M \in {\cal M}\}$ is a stable set of probability measures which are all  equivalent  martingale measures for $(S^k)_{1 \leq k \leq d}$.\\
 When the $M^i$ are continuous the preceding result is true without any restriction on $m$. 
\label{propStable}
 \end{proposition}

{\bf Proof}. From  Lemma 4.11 of \cite{BN03} ${\cal Q}({\cal M})$ is a stable set of probability measures equivalent to $P$.
 From the results on strongly orthogonal martingales \cite{P}, Chapter IV Section 3, it follows that for any $M \in {\cal M}$, ${\cal E}( M)$ is strongly orthogonal to $S^k$ for any $k$  and $Q_M$ is an equivalent  martingale measure for $(S^k)_{1 \leq k \leq d}$.\hfill $\square $\\

To construct TCPP extending the dynamics of reference assets $S^k$, one can take any stable subset of the set of equivalent local martingale measures for $S^k$,for example the set ${\cal Q}({\cal M})$ of Proposition \ref{propStable}, this defines a TCPP calibrated on the reference family. As soon as one adds options in the reference family, the threshold condition (\ref{eq_00bis}) has to be satisfied.
Notice that the set ${\cal M}_1$ introduced at the begining of Section \ref{sec:TCPP} is not stable in general. Our next goal is to construct a universal example of penalties which provides a convex TCPP calibrated on options. The technics used are those of right continuous  BMO martingales and the proof applies to any model $S^k$,  processes with jumps as well as processes in a Brownian filtration.\\  In~\cite{Sc} Schachermeyer introduced a notion of Robust No Arbitrage meaning that there is No Arbitrage with respect to a smaller bid ask spread.
In the same way we define here the notion of Robust No Free Lunch.
\begin{definition}
The reference family $((S^k)_{1 \leq k \leq d},(Y^l)_{1 \leq l \leq d})$ satisfies the Robust No Free Lunch Condition if there is $\epsilon>0$ such that 
is satisfies the No Free Lunch Condition when one replaces every  $C_{bid}(Y^l)$ (resp.  $C_{ask}(Y^l)$)  by  $C_{bid}(Y^l)+\epsilon$ (resp. $C_{ask}(Y^l)-\epsilon$) for any $l$ such that $C_{bid}(Y^l) \neq C_{ask}(Y^l)$.
\label{defRNFL}
\end{definition}
One  assumes in what follows, for simplicity, that for any $l$, $C^l_{bid} \neq C^l_{ask}$.\\
Denote ${\cal P}$ the predictable $\sigma$-algebra on $\R^+ \times \Omega$, and ${\cal B}(\R^j)$  the Borel $\sigma$-algebra on $\R^j$.
\begin{theorem}
Assume that the reference family $((S^k)_{0 \leq k \leq d},(Y^l)_{1 \leq l \leq d})$ satisfies the robust No Free Lunch condition with respect to the limit order books $C_{bid}(mY^l)_{m \leq M^l}$, $C_{ask}(nY^l)_{n \leq N^l}$. Denote $Q_0$ an equivalent local martingale measure for $S^k$ and $\epsilon>0$ such that for any $l$, 
 $C^l_{bid}+\epsilon < E_{Q_0}(Y^l)< C^l_{ask}-\epsilon$.  Let ${\cal M}$ be as in Proposition \ref{propStable}. Assume that $m < \frac{1}{16}$. 
 Let ${\cal Q}({\cal M})$ be the corresponding set  of equivalent probability measures $(Q_M)_{M \in {\cal M}}$  of Radon Nikodym derivative $\frac{dQ_M}{dQ_0}={\cal E}(M)$.
let $(b_i)_{1 \leq i \leq j}$, $b_i: \R^+ \times \Omega \times \R^j \rightarrow \R^+$ be non negative measurable maps with respect to the $\sigma$-algebra ${\cal P} \times {\cal B}(\R^j)$  such that $b_i(s,\omega,0,...,0)=0$. Assume that  there is a constant $B>0$ such that $b_i(s,\omega,x_1,...,x_j) \geq  B x_i^2$ for all $i$.  Denote $ b_i(s,H_1,...,H_j)$ the predictable process defined as $ b_i(s,H_1,...,H_j)(\omega)=b_i(s,\omega,H_{1,s}(\omega),...,H_{j,s}(\omega))$. For $M \in {\cal M}$ and  stopping times $0 \leq \sigma \leq \tau$ let
\begin{equation}
\alpha_{\sigma,\tau}(Q_M)=E_{Q_M}(\sum_{1 \leq i \leq j} \int\limits_{\sigma}^{\tau} b_i(s,H_1,...,H_j) d[M^i,M^i]_s|{\cal F}_{\sigma})
\label{eqb}
\end{equation}
Then
\begin{equation}\Pi_{\sigma,\tau}(X)=\esssup_{ Q_M \in {\cal Q}({\cal M})} (E_{Q_M}(X|{\cal F}_{\sigma})- \alpha_{\sigma,\tau}(Q_M))
\label{eqb2}
\end{equation}
defines a TCPP.
 Furthermore for $B$ large enough, The  TCPP  is calibrated on the reference family. Notice that the minimal acceptable $B$ depends only on $m$,  $\epsilon$, $\max(M^l,N^l)$ and $\max||Y^l||_{\infty}$ and not on the dynamics of $S^k$ nor on the set ${\cal M}$. 
\label{prop5}
\end{theorem}
The proof is given in the Appendix.
\begin{remark} 
 We get a similar result for weak calibration. 
\end{remark}

\section{Conclusion}
The motivation of this paper was to study and construct dynamic pricing procedures assigning to any financial instrument a dynamic limit order book in a arbitrage free way, extending  the dynamics of given basic assets and compatible with the observed limit order books  for  reference options.  This is done by making use of the theory of No Free Lunch TCPP introduced in \cite{BN05}.
 We have defined two notions of calibration for a Dynamic Pricing Procedure with respect to a reference family $((S^k)_{0 \leq k \leq d+1},(Y^l)_{1 \leq l \leq d})$ composed of two kinds of assets: the basic assets $(S^k)_{0 \leq k \leq d+1}$ for which the dynamic process is assumed to be known , and the assets $(Y^l)_{1 \leq l \leq d})$ (for example options) which are only  revealed at their maturity date (the stopping time $\tau_l$) and for which one  observes a limit order book at time $0$. One of the basic asset  $S^0$ is assumed to be strictly positive and is taken as num\'eraire. 
  The first notion of calibration, simply called calibration, assumes that the basic assets $(S^k)_{0 \leq k \leq d}$ are perfectly liquid.  A TCPP is said to be calibrated on the reference family  $((S^k)_{0 \leq k \leq d},(Y^l)_{1 \leq l \leq p})$  and the limit order books $C_{bid}(mY^l)_{m \leq M^l}$, $C_{ask}(nY^l)_{n \leq N^l}$  if it extends the dynamics of the basic assets $(S^k)_{0 \leq k \leq d}$ and it is compatible with the observed limit order books for the options $(Y^l)_{1 \leq l \leq p}$. The second notion  called weak calibration takes into account the limit order books  associated with the basic assets. 
We have characterized TCPP calibrated or weakly calibrated on the 
reference family $((S^k)_{0 \leq k \leq d},(Y^l)_{1 \leq l \leq p})$  and the limit order books  in terms of their dual representation. In case of calibration, any probability measure in the dual representation of the TCPP has to be an equivalent local martingale measure for the process $(S^k)_{0 \leq k \leq d}$ while in case of weak calibration this is only the case for probability measures with zero  penalty. In both cases there is a threshold condition on the penalty.\\
 We have extended to that context the notion of No Free Lunch, replacing the usual notion of dynamic strategy with respect to the basic assets $(S^k)_k$ by the sum of a dynamic strategy with respect to the basic assets $S^k$ and of a static strategy with respect to the options $Y^l$. 
We have proved the following generalization of  Kreps-Yan Theorem: there is No Free Lunch with respect to the reference family $((S^k)_{0 \leq k \leq d},(Y^l)_{1 \leq l \leq p})$  and the limit order books $C_{bid}(mY^l)_{m \leq M^l}$, $C_{ask}(nY^l)_{n \leq N^l}$  if and only if  there is an equivalent local martingale measure $Q$ for the process $(S^k)_{0 \leq k \leq d}$ such that, for every $l$, and any $n \geq 0$, $C_{bid}(nY^l) \leq E_Q(nY^l) \leq C_{ask}(nY^l)$. Furthermore, the No Free Lunch condition is also equivalent to the existence of a No Free Lunch TCPP calibrated on the reference family.\\
We have illustrated our results with two examples: The first one is the case of TCPP calibrated on very liquid options $C_{bid}(nY)=C_{ask}(nY^l)= n C^l\;\; \forall n \in \N$. We have proved in that case that the process $(Z^l)_t=E_Q(Y^l|{\cal F}_t)$ is independent on the probability measure $Q$ involved in the dual representation of the TCPP. Therefore the options can be used to hedge (dynamically) as well as the basic assets. The second example is that of a stochastic volatility model.\\
 We have also used the powerful technique of right continuous BMO martingales in order to prove the existence of convex (not sublinear)  No Free Lunch TCPP calibrated on the reference family.    We have produced a generic construction of a  convex No Free Lunch TCPP calibrated on the reference family  $((S^k)_{0 \leq k \leq d},(Y^l)_{1 \leq l \leq p})$  and the limit order books $C_{bid}(mY^l)_{m \leq M^l}$, $C_{ask}(nY^l)_{n \leq N^l}$ as soon as this reference family satisfies  the Robust No Free Lunch condition.  Such a No Free Lunch TCPP is constructed inside the new family first introduced in \cite{BN03}. This construction is made in a very general setting of locally bounded stochastic processes for which  jumps are allowed.\\ 
The advantage of dynamic pricing making use of  TCPP  is that it not only takes into account the liquidity risk and the properties of the limit order books, but also it induces more robustness in the prices. A small variation  in  the values of the limit order books of the options on which the TCPP is calibrated induces only a small modification of the constructed TCPP.


\renewcommand{\theequation}{A-\arabic{equation}} 
\setcounter{equation}{0}  
\renewcommand{\thesection}{A} 
\setcounter{section}{0}  

\section{Appendix}
\label{sec:Appendix} 
\subsection{Proof of the extended First Fundamental Theorem}
\label{secEFFT}
We prove directly the extended version formulated in Theorem \ref{thm2}, Section \ref{sec:EV}. It gives also a proof of Theorem \ref{thm1} of Section \ref{sec:fund. Thm}.\\
{\bf Proof}.
We begin with the easiest implications.\\
- {\it ii}) implies {\it iii}): We  define the TCPP $\Pi$ as follows:
for any stopping times $\sigma \leq \tau$, 
$\Pi_{\sigma,\tau}(X)=E_R(X|{\cal F}_{\sigma})$.
 As  $C_{bid}(mY^l) \leq E_R(m Y^l)$ for $m \leq M^l$, and  $ E_R(n Y^l) \leq C_{ask}(nY^l)$ for any $n \leq N^l$, it follows that $\Pi$ is calibrated on the reference family. It has No Free Lunch  as $R$ is equivalent to $P$ and the penalty associated with $R$ is equal to $0$.\\
- {\it iii}) implies {\it iv}) is trivial.\\
- {\it iv}) implies {\it i}): \\ 
-Let $\Pi$ be a No Free Lunch TCPP weakly calibrated on the reference family. Consider its dual representation, equation (\ref{eq_NFL2}). As the TCPP $\Pi$ has No Free Lunch, the set ${\cal M}^0$ of equivalent probability measures with zero minimal penalty (equation(\ref{eqzeropen})) is non empty. Define
$$(\Pi^0)_{0,\infty}(X)=\sup_{Q \in {\cal M}^0}E_Q(X)$$
The sets $\overline C$ and $\tilde K$ are those defined in Section \ref{sec:fund. Thm}. Prove that $\forall X \in \overline C$,
 $(\Pi^0)_{0,{\infty}}(X) \leq 0$. Let $Z \in \tilde K$ \\
$Z=\sum_{i=1}^{n} \sum_{k=1}^{d} (h^k)_i (S^k_{\sigma_i}-S^k_{\sigma_{i-1}})+
\sum_{l=1}^{p}(\gamma^l-\beta^l)Y^l+(\gamma^0-\beta^0)-g$ \\
for some $g \in  L^{\infty}_+(\Omega,{\cal F},P)$.  From Theorem \ref{thm2.3}, any $Q$ in ${\cal M}^0$ is an equivalent local martingale measure for the process $S^k$. Thus 
$E_Q(Z)=\sum_{l=1}^{p}(\gamma^l-\beta^l)E_Q(Y^l)+(\gamma^0-\beta^0)-E_Q(g)$. As $\Pi_{\sigma,\tau}$ is weakly calibrated on the reference family, the minimal penalty satisfies  equation (\ref{eq_00}) of Theorem \ref{thm2.3}. For $Q \in {\cal M}^0$, $\alpha^m_{0,\infty}(Q)=0$ so  
$$E_Q(Z) \leq \sum_{l=1}^{p}( C_{ask}(\gamma^l Y^l)- C_{bid}(\beta^lY^l))+(\gamma^0-\beta^0) \;\leq 0.$$ 
 As $\overline C$ is the weak* closure of the cone generated by $\tilde K$, it follows that for every 
$X \in \overline C$, and  $Q$ in ${\cal M}^0$,  $ E_Q(X) \leq 0$.
And then  $(\Pi^0)_{0,{\infty}}(X) \leq 0$.\\
 - Assume now that $X\in \overline C \INTER L^{\infty}_+(\Omega,{\cal F},P)$. 
 $X \geq 0$. If $X\neq 0$ in $L^{\infty}$, there is $\alpha \in \R^*_+$ and $A$ with $P(A)>0$ such that $X \geq \alpha 1_A$. Let $Q \in {\cal M}^0$. $Q$ is equivalent with $P$, so $(\Pi^0)_{0,{\infty}}( X) \geq \alpha Q(A)> 0$. Thus we get a contradiction. 
So  $\overline C \INTER L^{\infty}_+(\Omega,{\cal F},P)=\{0\}$.\\
- {\it i}) implies {\it ii}):
The proof follows that of Theorem 5.2.2. of~\cite{DS02}.
 Let $f \in  L^{\infty}_+(\Omega,{\cal F},P)$. As $\overline C$ is closed for the weak * topology, and 
$\{f\}$ is compact, from Hahn Banach Theorem, there is $g \in L^1$, $g \neq 0$, such that 
$$sup_{Z \in \overline C} E(gZ) < E(fg)$$ 
As $\overline C$ is a cone, $sup_{Z \in \overline C} E(gZ)=0$ and $0 < E(fg)$. We have
 $- L^{\infty}_+ \subset \overline C$ so $g \geq 0$. 
 The exhaustion argument of  the proof of Theorem 5.2.2. of~\cite{DS02}  applies without any change. So we get $g_0$ strictly positive $P$ a. s. such that $sup_{Z \in \overline C} E(g_0Z)=0$.
Denote $Q$ the probability measure whose Radon Nikodym derivative is $\frac{g_0}{E(g_0)}$. 
$\{(H.S)_{\infty}\}$ where $H$ is an admissible simple strategy is a vector space contained in $\overline C$.  The linear form $E_Q$ is non positive on this vector space. So it has to be identically equal to $0$ on it.\\
 It follows then from Lemma 5.1.3. of~\cite{DS02} that $S$ is a local martingale under $Q$.
 Let $l\in \{1...p\}$ and $m \leq M^l$. As $-mY^{l}+C_{bid}(mY^l)\in K$ ($\beta^{l}=m\;\gamma^0=C_{bid}(mY^l)$) it follows that
 $C_{bid}(mY^l)-E_Q(mY^{l}) \leq 0$. In the same way  $E_Q(nY^{l}) \leq C_{ask}(nY^l)$ for $n \leq N^l$ (as $nY^{l}-C_{ask}(nY^l)\in K$).\hfill $\square $
\subsection{Proof of Theorem \ref{prop5}}
\label{secThm}
Before starting the proof of the theorem we prove two lemmas.
\begin{lemma}
Assume that $M$ is a $Q_0$-martingale of BMO norm less than $m$, $m<\frac{1}{16}$.
For any stopping time $T$, 

\begin{equation}
|1-{\cal E}( M)_T| \leq |M_T|+[M,M]_T)exp(|M|_T+[M,M]_T)
\label{eq17}
\end{equation}

\label{lemma4.6}
\end{lemma}
{\bf Proof}.
Recall (\cite{P}) that 
$${\cal E}(M) _T=exp(M_T-\frac{1}{2}([M,M]^c)_T) \Pi_{s \leq T}(1+\Delta M_s)e^{-\Delta M_s}$$
As $m<1$, each term of the product is positive and less than $1$, therefore
\begin{eqnarray}
{\cal E}(M) _T-1 & \leq & exp(M_T-\frac{1}{2}([M,M]^c)_T)-1 \nonumber \\
 & \leq & exp(|M_T|)-1 \leq |M_T|exp(|M_T|)
\label{eq18}
\end{eqnarray}
Apply the inequality $\frac{1+x}{e^x} \geq e^{-x^2} $, for $|x|<\frac{1}{16}$, (cf \cite{DDM01})with $x = \Delta M_s$ 
\begin{equation}
{\cal E}(M) _T  \geq  exp(-|M_T|-[M,M]_T)
\label{eq19}
\end{equation}
Therefore 
\begin{equation}
 1-{\cal E}(M) _T  \leq (|M_T|+[M,M]_T) 
\label{eq20}
\end{equation}
Lemma \ref{lemma4.6} follows from the equations (\ref{eq18}) and (\ref{eq20}).
\begin{lemma}
Let $m <\frac{1}{16}$. There is a constant $K$ and an integer $r>0$ depending only on $m$ such that for any  $Q_0$-martingale $M$ of BMO norm less than $m$, for any  stopping time $T$,

\begin{equation}
 E_{Q_0}(|1-{\cal E}(M)_T|) \leq K ((E_{Q_0}([M,M]_T))^{\frac{1}{r}}
\label{eq21}
\end{equation}

\label{lemma 4.7}
\end{lemma}

{\bf Proof}.
Choose $q$ a positive  integer $q > \frac {1}{1-16m}$. let $p \in \R^+$  such that $\frac {1}{p}+\frac {1}{q}=1$, then $16pm<1$. In all the following $E$ means $E_{Q_0}$.
From Lemma \ref{lemma4.6} and H\"older inequality, it follows that
\begin{equation}
E(|{\cal E}(M_T)-1|)  \leq  \{E(|M|_T^q)^{\frac{1}{q}}+E([M,M]_T^q)^{\frac{1}{q}}\} \{E(exp(p|M|_T+p[M,M]_T))\}^{\frac{1}{p}}
\label{eq21b}
\end{equation}
 Applying the Cauchy Schwartz inequality,
\begin{equation}
 E(exp(p|M|_T+p[M,M]_T)) \leq \{E(exp(2p|M|_T))\}^{\frac{1}{2}}\{E(exp(2p[M,M]_T))\}^{\frac{1}{2}}
\label{eq21t}
\end{equation}
Applying John Nirenberg inequality and Lemma 1 of \cite{DDM01}, it follows from (\ref{eq21t}) that

\begin{equation}
 E(exp(p|M|_T+p[M,M]_T)) \leq (\frac{1}{(1-16pm)})^{\frac{1}{2}} (\frac{1}{(1-2pm^2)})^{\frac{1}{2}}
\label{eq21q}
\end{equation} 

On the other hand, again from Cauchy Schwartz inequality,
\begin{eqnarray}
 E((|M|_T)^q)^{\frac{1}{q}} \leq  \{E([M,M]_T)\}^\frac{1}{2q} \{E(|M|_T^{2q-2})\}^{\frac{1}{2q}}\\
  E(([M,M]_T)^q)^{\frac{1}{q}} \leq  \{E([M,M]_T\}^\frac{1}{2q}) \{E([M,M]_T^{2q-1})\}^{\frac{1}{2q}}
\label{eq21w}
\end{eqnarray} 

From Burkholder Davis Gundy inequality, (Theorem 30 of~\cite{Meyer},  there is a constant $c$ such that
$$ E(|M|_T^{2q-2}) \leq c  E([M,M]_T^{q-1})$$
and for every integer $n$, it follows from the proof of Lemma 1 of \cite{DDM01}
that 
\begin{equation}
E([M,M]_T^{n}) \leq m^{2n} n!
\label{eq21r}
\end{equation}
 
This proves the lemma with $r=2q$.

\begin{lemma}
There is a constant $K_1$ depending only on $m$ such that for any  $Q_0$-martingale $M$ of BMO norm less than $m$, for any stopping time $T$, 
\begin{equation}
 E([M,M]_T) \leq K_1 (E({\cal E}(M) _T [M,M]_T))^{\frac{1}{2}}
\label{eq22}
\end{equation}
\label{lemma 4.8}
\end{lemma}
{\bf Proof}.
From Cauchy Schwartz inequality, and then H\"older inequality with $p$ such that $16pm<1$ $q \in\N^*$, and $\frac{1}{p}+\frac{1}{q}=1$, as in the preceding proof, 
\begin{equation}
 E([M,M]_T) \leq  \{E({\cal E}(M)_T [M,M]_T)\}^{\frac{1}{2}}  \{E([M,M]_T^q\}^{\frac{1}{2q}}  \{E({\cal E}(M)_T^{-p})\}^{\frac{1}{2p}}
\label{eq35}
\end{equation}
From equation (\ref{eq21r}) 
$ E(([M,M]_T)^q) \leq m^{2q} q!$
From inequalities (\ref{eq19}) and (\ref{eq21q}), 
$$ E(({\cal E}(M)_T)^{-p})\leq  (\frac{1}{(1-16pm)} \frac{1}{(1-2pm^2)})^{\frac{1}{2}}$$
This proves equation (\ref{eq22}).
\paragraph{{\bf Proof of Theorem \ref{prop5}.}}
 From proposition 5 of~\cite{BN04}, we already know that equation (\ref{eqb2}) with (\ref{eqb})  defines a TCPP. $\alpha_{\sigma,\tau}(Q_0)=0$ for every $\sigma \leq \tau$, thus the TCPP has No Free Lunch. Furthermore, from Proposition \ref{propStable},  for any $M$, $Q_M$ is an equivalent martingale measure for every $S^k$.
 So from proposition \ref{propStable1}, we just have to find a condition on $B$ such that  the threshold condition (\ref{eq_00bis}) is satisfied  for any stopping time $T$, and any probability measure $Q_M$ in ${\cal Q}({\cal M})$.
By hypothesis on $b_i$, for any $M$, for any stopping time $T$
$\alpha_{0,T}(Q_M) \geq B E_{Q_M}([M,M]_T)$. 
Thus it is enough to verify that 
\begin{eqnarray}
B E_{Q_0}({\cal E}(M)_{T}[M,M]_T) \geq  \sup_{ \{l\;|\; \tau_l \leq T\} } (\sup_{m \leq M^l}((C_{bid}(mY^l)-E_{Q_M}(mY^l)),\nonumber \\ \sup_{n \leq N^l}(E_{Q_M}(nY^l)-C_{ask}(nY^l))
\label{eq_8c}
\end{eqnarray}
Choose $0< \epsilon \leq inf_{\{1 \leq l \leq p\}}( C_{ask}(Y^l)- E_{Q_0}(Y^l), E_{Q_0}(Y^l)-C_{bid}(Y^l))$ , so for any $m \leq M^l$, $E_{Q_0}(mY^l)-C_{bid}(m Y^l) \leq m \epsilon$. (idem for $C_{ask}(nY^l)$. Thus to satisfy  (\ref{eq_8c}), it is sufficient that for any $n \leq \sup(M^l,N^l)$
\begin{equation}
B E({\cal E}(M)_{T}[M,M]_T)  +n \epsilon \geq \sup_{ \{l\;|\; \tau_l \leq T\} } n|(E_{Q_0}(Y^l)-E_{Q_M}(Y^l)| 
\label{eq_8d}
\end{equation}
Notice that $ (E_{Q_0}(Y^l)-E_{Q_M}(Y^l)=E_{Q_0}((1-{\cal E}(M)_{T})Y^l)$.

From Lemma \ref{lemma 4.7}, and Lemma \ref{lemma 4.8}, there is $\tilde K$ depending only on $m$ and $\epsilon$ such that for any $l$, 
\begin{equation}
|E({\cal E}(M_T)-1)Y^l)| \leq  \tilde K ||Y^l||_{\infty}(E({\cal E}(M) _T [M,M]_T))^{\frac{1}{2r}}
\label{eq30}
\end{equation}
There is a constant $B_0>0$ such that for any $x>0$, 
$$ \tilde K x^{\frac{1}{2r}} (\max(||Y^l||_{\infty})\leq B_0 x+\epsilon$$
Then $B \geq \max(M^l,N^l)B_0$ satisfies the required conditions.
\hfill $\square $

\end{document}